# Unusual thermal Hall effect in the 3$d$ cobalt Kitaev system Na$_2$Co$_2$TeO$_6$

Heejun Yang[1,2,*], Chaebin Kim[1,2,*], Ysun Choi[1,2], Jun Han Lee[3], Gaoting Lin[4], Jie Ma[4], Marie Kratochvílová[5], Petr Proschek[5], Eun-Gook Moon[6], Ki Hoon Lee[7], Yoon Seok Oh[3,#] and Je-Geun Park[1,2,$]

[1]Center for Quantum Materials, Seoul National University, Seoul 08826, Republic of Korea

[2]Department of Physics and Astronomy, Seoul National University, Seoul 08826, Republic of Korea

[3]Department of Physics, Ulsan National Institute of Science and Technology, Ulsan 44919, Republic of Korea

[4]Key Laboratory of Artificial Structures and Quantum Control, School of Physics and Astronomy, Shanghai Jiao Tong University, Shanghai 200240, China

[5]Charles University, Faculty of Mathematics and Physics, Department of Condensed Matter Physics, Ke Karlovu 5, 121 16 Prague 2, Czech Republic

[6]Department of Physics, Korea Advanced Institutes of Science and Technology, Daejeon 34141, Republic of Korea

[7]Department of Physics, Incheon National University, Incheon 22012, Republic of Korea

[*]These authors contributed equally to this work

[#]ysoh@unist.ac.kr

[$]jgpark10@snu.ac.kr

**Abstract**

Kitaev physics has recently attracted attention in condensed matter for its anticipated novel quantum spin liquid state. The thermal transport measurement is crucial for probing the novel features of charge-neutral quasiparticles. In this letter, we report a significant thermal Hall effect in Na$_2$Co$_2$TeO$_6$ (NCTO), a Kitaev quantum spin liquid candidate, when the magnetic field is applied along the out-of-plane direction of the honeycomb plane. The thermal conductivity ($\kappa_{xx}$) and thermal Hall conductivity ($\kappa_{xy}$) in NCTO reveal distinct magnetic field dependences below and above the Neel temperature ($T_N$) of 27 K. For $T>T_N$, $\kappa_{xx}$ has a monotonic decrease in the field dependence, while $\kappa_{xy}$ persists up to $T^*$ = 150 K. On the other hand, both $\kappa_{xx}$ and $\kappa_{xy}$ exhibit complex field dependence for $T<T_N$.

Quantum entanglement is the most intriguing concept and has become the essential keyword in the new era of quantum information science. For several decades, it has been hypothesized that this fascinating phenomenon can be realized in certain magnetic materials. One of them is quantum spin liquid (QSL) which was conjectured to possess long-range quantum entanglement without any long-range magnetic order [1]. It has ever been a long-sought issue for both theoretical and experimental development. Recently, the Kitaev model, an exactly solvable 2D model Hamiltonian, has been theoretically proposed to exhibit a new QSL ground state with bond-dependent Ising interactions [2]. Moreover, this Kitaev's QSL (KQSL) state is expected to host itinerant Majorana fermions and topologically non-trivial $Z_2$ fluxes.

Theoretical and experimental breakthroughs [3–9] have led to the discovery of Kitaev candidate materials with potential KQSL. However, non-negligible Heisenberg interactions and other off-diagonal anisotropy are found to prevent them from achieving a genuine KQSL state without external magnetic fields. Among such rare examples, α-RuCl₃ has been the best-known case, probably closest to the KQSL state with an in-plane magnetic field at $μ_0H_c$ = 8~10 T [10–15]. Although α-RuCl₃ is believed to have KQSL under the in-plane magnetic fields, a noticeable change has never been observed for the out-of-plane magnetic fields up to 14 T [16]. Note that all the reported materials of KQSL have ferromagnetic Kitaev interaction [4,17,18]. Nonetheless, a recent theoretical study suggested that antiferromagnetic Kitaev interaction could yield intermediate gapless QSL [19]. Given this new theoretical insight and the limitations in currently available materials, finding another Kitaev candidate material is an urgent issue in the field. Furthermore, it would be much more interesting to find materials with antiferromagnetic Kitaev interactions as such a case has never been reported before.

Research interest has recently increased in the 3*d* cobalt (Co) honeycomb system. It was even proposed that Co systems may well realize the Kitaev model despite its relatively weak spin-orbit coupling compared to 4*d* or 5*d* compounds [20,21]. A spin-orbital entangled $J_{eff}$=1/2 state of the $Co^{2+}$ ions [22] could satisfy an essential requirement for realizing Kitaev interaction [23–25]. Interestingly, it was reported that cobalt honeycomb compound $Na_2Co_2TeO_6$ (NCTO) has significant Kitaev interaction, even larger than other non-Kitaev terms [26,27]. More importantly, it is expected to have antiferromagnetic Kitaev interaction based on the inelastic neutron scattering experiments and subsequent analysis [27]. NCTO has a zig-zag order at $T_N$ = 27 K with a propagation vector of Q = (0.5, 0, 0) [20,26–33]. The identical bond length for the nearest neighbor leads to the ideal honeycomb lattice of $Co^{2+}$ (space group P6₃22), making it more likely to have the isotropic Kitaev interaction (see Fig. 1(a)). Another interesting point is that NCTO has a field-induced spin-disordered state similar to α-RuCl₃ when the magnetic field is applied along the *ab*-plane [32].

Thermal transport measurement has been widely used to study low-energy quasiparticle excitations in a perfect insulator, including the QSL system. Especially the thermal Hall effect, a thermal version of the electrical Hall effect, has been shown to reveal the topological nature of quasiparticles [34–39]. Here, we investigated the nature of the KQSL candidate in NCTO by measuring the longitudinal thermal conductivity and thermal Hall coefficient.

In this letter, we report the intriguing magnetic field dependence of in-plane thermal conductivity ($\kappa_{xx}$) and thermal Hall coefficient ($\kappa_{xy}$) in NCTO. In the supposedly paramagnetic state, $\kappa_{xx}(H)$ shows pronounced temperature effects on the magnetic field dependence while $\kappa_{xy}(H)$ exhibits negative field dependence: which is similar to the field dependence of $\Delta\kappa_{xx}(H)/\kappa_{xx}(0)$ and $\kappa_{xy}(H)$ of α-RuCl₃ [38]. In the zig-zag antiferromagnetic phase, both $\kappa_{xx}(H)$ and $\kappa_{xy}(H)$ could be described by a complex function of magnetic field, including sign-changing behavior of $\kappa_{xy}$ at low temperature. We provide balanced discussions about the candidates of this unusual thermal Hall effect, such as phonon, magnon, magnon-phonon coupling, and other non-trivial spin excitation. Lastly, we discuss the resemblance between NCTO and α-RuCl₃ based on our observations. This indicates that these two KQSL candidates might share a similar origin for the unusual thermal Hall effect.

High-quality single crystals of NCTO were grown by a flux method using a recipe modified from Ref. [28]. The NCTO shows a representative antiferromagnetic anomaly in the magnetization and the specific heat measurement at $T_N$ = 27 K, as shown in Fig. 1(b) and 1(c) [28]. The specific heat data were collected from 2 to 40 K under various magnetic fields up to 19.3 T using a 20-T cryostat (Cryogenic Ltd.). Fig. 1(c) shows specific heat results taken under the magnetic fields along the *c*-axis, where a sharp peak at $T_N$ broadens with the magnetic field, with $T_N$ being suppressed by 5 K at 19.3 T. The modest decrease in $T_N$ indicates that the long-range magnetic order gets more slowly suppressed than in a recent study of in-plane field dependence on the same compound [32]. In an ideal Kitaev model, the KQSL state should be stabilized for any arbitrary field direction [19]. Considering that α-RuCl₃ shows no noticeable change of $T_N$ under the magnetic field along the out-of-plane, NCTO could be seen as a better candidate for the KQSL than α-RuCl₃.

For the thermal Hall measurement, a single crystal NCTO sample was prepared in a plate-like rectangular shape with a dimension of 5.17×2.07×0.054 mm³. The standard steady-state method was employed for the thermal transport experiment in a high vacuum, as shown in Fig. 1(d). The chip resistance heater induced heat current $J_Q$ along the crystallographic *a*-axis (*a* // x) and three thermometers simultaneously probed longitudinal ($\Delta T_x$) and transverse ($\Delta T_y$) temperature differences under the applied magnetic field (***H***-field) along the *c*-axis (*c* // z). Silver wires were used for thermal contacts between the NCTO sample and the heater/thermometers. We employed SrTiO₃ capacitors as thermometers to avoid the strong magnetic field effect and self-heating and performed in-situ calibrations [40]. $\kappa_{xx}(H)$ was measured with a slowly varying field sweep. During the thermal Hall measurement, a constant magnetic field was applied in a step-by-step mode under the iso-thermal condition, and every data points were taken by averaging $\Delta T_y$ for a few minutes to minimize magneto-caloric effect. And $\Delta T_y$ was collected from both ramping directions of magnetic field (ramping up and down) to correct for possible hysteretic behavior affected by a weak ferrimagnetic moment along the *c*-axis. Using the anti-symmetrization technique in Ref. [11], we compensated for the longitudinal contamination in $\Delta T_y$ due to inevitable experimental misalignment of the contacts and possible hystereic longitudinal $\Delta T_x$.

Fig. 2(a) represents $\kappa_{xx}$ of NCTO as a function of temperature. In the absence of the magnetic field, $\kappa_{xx}$ shows a peak near 50 K and decreases as the temperature increases, similar to the recent report of thermal conductivity [31]. The sharp peak in $\frac{d\kappa_{xx}}{dT}$ is clearly observed near $T_N$ ~ 26.3 K (see inset of Fig. 2(a)). Under the magnetic field of 14 T along the *c*-axis, $\kappa_{xx}(T)$ still maintains the overall shape but exhibits a slight suppression around the peak position. Fig. 2(b) and 2(c) summarize the developments of $\kappa_{xx}(H)$ as a function of the magnetic field below and above $T_N$. For *T*>*T*$_N$, monotonic negative field dependence in $\kappa_{xx}(H)$ persists up to *T*\* = 150 K, as shown in Fig. 2(b). In a typical magnetic insulator, phonon ($\kappa_{xx}^{ph}$) and spin ($\kappa_{xx}^{spin}$) terms can contributes to $\kappa_{xx}$: $\kappa_{xx} = \kappa_{xx}^{ph} + \kappa_{xx}^{spin}$. Assuming phonon heat transport dominant in the paramagnetic state [31], $\kappa_{xx}^{ph}$ is expected to increase with magnetic fields since spin alignment due to the applied magnetic field reduces the spin-phonon scattering [36,37,41–44]. Thus, the clear negative field dependence of $\kappa_{xx}(H)$ in NCTO for *T*>*T*$_N$ could be interpreted into the following ways: some heat carrier associated with spin excitations ($\kappa_{xx}^{spin}$) directly contributes to decreasing $\kappa_{xx}$ [43], or the spin excitation indirectly decreases $\kappa_{xx}^{ph}$ via complex spin-phonon interactions. On the other hand, for $T<T_N$, $\kappa_{xx}(H)$ (Fig. 2(c)) exhibits a non-monotonic field dependence with weak hysteresis [45].

Fig. 3(a) shows thermal Hall conductivity divided by temperature ($\kappa_{xy}/T$) in the paramagnetic state, where the negative signed sub-linear curves are repeatedly observed up to $T$ = 110 K. This sub-linear curve at $T<T_N$ starts to be modified in the low field region ($\mu_0 H$ < 9 T), as shown in Fig. 3(b). Such modification gets more pronounced as the temperature lowers, which leads to complex sign-changing behavior in $\kappa_{xy}/T$. Overall temperature dependence of $\kappa_{xy}/T$ is summarized in Fig. 3(c) at a constant magnetic field of 14 T. A small negative value of $\kappa_{xy}/T$ is recorded at 200 K and grows negatively as the temperature decreases. $\kappa_{xy}/T$ reaches the lowest value at $T_N$ and then approaches toward zero with negative values.

Before getting to the discussion, we should note that the origin of the thermal Hall effect on KQSL candidates is still controversial, even for α-RuCl$_3$ [46], the most researched KQSL candidate. Therefore, we would like to open to all the possibilities for discussion before giving still speculative explanations of our experimental observations.

First of all, we observed that the temperature dependence of $\kappa_{xy}/T$ at 14 T in Fig. 3(c) can be fitted using a phenomenological model based on a generic model of Berry curvature in the quasiparticle band [47]. As can be seen in the figure, $A_0 \exp(-T/T_0) + C_0$ shows excellent agreement with the experimental $\kappa_{xy}/T$ for $T>T_N$, which indicates that the nonzero thermal Hall effect in the paramagnetic state of NCTO originates most likely from the nontrivial Berry curvature of the quasiparticle bands. On the other hand, we can consider another function for the temperature dependence of $\kappa_{xy}/T$ below $T_N$. For example, $A_1 \exp(-\Delta/T)$ represents the thermal Hall effect from gapped quasiparticles with a fitting parameter $\Delta$ for an effective activation gap [47]. Another type of function like $A_2 T^2$ is expected for the acoustic phonon Hall effect in magnetic solid [48]. We tried both fitting functions to explain our $\kappa_{xy}/T$ taken at 14 T and, as shown in the inset of Fig. 3(c), the $A_1 \exp(-\Delta/T)$ function gives better fitting results, in favor of the gapped nature of quasiparticle band. However, we also note that $\kappa_{xy}/T$ displays sign changing behavior at lower fields, which makes it problematic to use the same function for the full field range at low temperatures (Fig. S3(a)). Thus, we think that one needs thermal Hall measurement data taken at further lower temperature in order to have a fuller picture of the possible quasiparticles.

For $T<T_N$, magnons could be a natural candidate for $\kappa_{xy}$, and it is known that the generalized Kitaev-Heisenberg model can produce non-zero magnon Hall conductivity [49]. To check this possibility, we made a theoretical estimation of the magnon Hall effect ($\kappa_{xy}^{mag}$) in NCTO by using linear spin wave theory (LSWT) at temperature of 0.3$T_N$ with the reported magnetic Hamiltonian for NCTO [26,27,30,32,50]. However, we find that all the calculated $\kappa_{xy}^{mag}$ has thermal Hall conductivity with two orders of magnitude smaller than our experimental values (see Fig. S4). But, it is to be note d that despite this discrepancy one has to be careful before ruling out the magnon Hall scenario for the following reason: the reported magnetic Hamiltonian is not complete yet, not capturing the single crystal inelastic neutron scattering data well [51]. Moreover, LSWT might not be enough to describe the effective spin-1/2 system because of quantum fluctuations. The spin wave theory with Schwinger boson representation [42,43,52,53] and an accurate model Hamiltonian would be desirable to check the magnon scenario in future study.

On the other hand, one can expect significant phonon contribution in $\kappa_{xy}$ since several huge phonon-mediated Hall effects are reported so far [54–58], even in the non-magnetic insulator SrTiO$_3$ [54]. Unfortunately, we found that the mechanisms proposed for SrTiO$_3$ are difficult to be applied to NCTO since it requires structural domains with huge dielectric constant ($\epsilon \approx 2 \times 10^4$) [59] or quantum paraelectric phase [60], both of which are absent in NCTO [61]. Nevertheless, in principle, phonon can still contribute to non-zero $\kappa_{xy}$, since recent theoretical studies showed that non-trivial Berry curvature can be induced by coupling between phonon and original spin Hamiltonian [62–64]. Indeed, such spin-phonon coupled Hamiltonian provides quite good explanation for anomalous magnon Hall effect in VI$_3$ at low temperature range, which shows large deviation from magnon only model [65]. Hence, phonon scenario could not yet be discarded explicitly for NCTO.

For $T>T_N$, $\kappa_{xy}(T)$ diminishes slowly up to 200 K (Fig. 3(c)). Related to this discussion, our magnetic-specific heat measurement on NCTO up to 150 K (Fig. S5) found a broad peak around 100 K in addition to a sharp peak of antiferromagnetic phase transition. This observation implies a two-step release of magnetic entropy and non-trivial spin excitations in NCTO at high temperatures, similar to α-RuCl$_3$ [13]. Having said that, we should also consider the phonon effect coupled with this non-trivial spin degree of freedom [66]. Surprisingly, recent thermal Hall measurement in α-RuCl$_3$ represents that $\kappa_{xy}$ in α-RuCl$_3$ possesses one feature of phonon-mediated Hall effect: similar temperature dependence between $\kappa_{xx}(T)$ and $\kappa_{xy}(T)$ [67]. Moreover, a recent study confirmed the phonon renormalization and the coupling between phonon and non-trivial spin excitation using inelastic X-ray scattering in α-RuCl$_3$ [68]. As shown in Fig. S6, NCTO also shows similar $\kappa_{xx}(T)$ and $\kappa_{xy}(T)$ with a broad peakand comparable peak positions. Detailed future studies of phonon-based scenario would be necessary for this puzzling phenomenon.

Even though we cannot take one's side among those suggested possibilities, we want to stress several similarities in thermal transport properties between NCTO and α-RuCl$_3$. As shown in the recent study, the in-plane $\kappa_{xx}(H)$ with in-plane magnetic field ($J_Q$ // $H$ // $ab$-plane) in NCTO is very similar to α-RuCl$_3$, indicative of possible Kitaev-Heisenberg paramagnon present in NCTO [31,69]. We also noted other similarities between NCTO and α-RuCl$_3$ with a magnetic field along the $c$-axis. For $T>T_N$, monotonic decreasing $\kappa_{xx}$ in Fig. 2(b) was also observed in α-RuCl$_3$ up to temperatures much higher than $T_N$ with a comparable decreasing rate [38]. Interestingly, both monotonic field dependence (Fig. 3(a)) and exponential decaying temperature dependence (Fig. 3(c)) of $\kappa_{xy}/T$ also indicate similar thermodynamics of $\kappa_{xy}$ at work for both NCTO and α-RuCl$_3$ [38,67,70]. Therefore, we would like to conclude that both NCTO and α-RuCl$_3$ share the same origin for the thermal Hall effect. However, it is worthwhile to note here that several competing scenarios have since been put forward for the thermal Hall effect even in α-RuCl$_3$: for instance, Majorana fermion [71–76], topological magnon [77,78], and spin-phonon coupling [67].

In summary, we measured specific heat and thermal conductivity tensor of NCTO with magnetic fields along the $c$-axis, which is a promising candidate for an exotic magnetic ground state of Kitaev physics. Specific heat measurements under various constant magnetic fields show that the antiferromagnetic order gets slowly suppressed by increasing the magnetic field up to 19.3 T. In contrast, it is found that both in-plane thermal conductivity ($\kappa_{xx}$) and thermal Hall conductivity ($\kappa_{xy}$) in NCTO are sensitive to magnetic field up to much higher temperatures than $T_N$, showing temperature and field dependences similar to α-RuCl$_3$, except for the opposite signs of $\kappa_{xy}$. Thus, it is plausible that some gapped quasiparticles can produce such thermodynamic properties in our experimental results. However, further studies are necessary to shed light on the nature of quasiparticles, although overall similarities between NCTO and α-RuCl$_3$ may as well imply a common mechanism for the unusual thermal Hall effect.

**Acknowledgment**

The work at CQM and SNU was supported by the Leading Researcher Program of Korea's National Research Foundation (Grant No. 2020R1A3B2079375). J.H.L. and Y.S.O. acknowledge support from the Basic Science Research Programs through the National Research Foundation of Korea (NRF) (NRF-2020R1A2C1009537). High-field specific heat experiments were performed in MGML, supported by the Czech Research Infrastructures program (project no. LM2018096). G.T.L and J.M. thank the financial support from the National Science Foundation of China (Nos. U2032213 and 11774223).

# **Reference**

[1] P. W. Anderson, Science (80-. ). **235**, 1196 (1987).

[2] A. Kitaev, Ann. Phys. (N. Y). **321**, 2 (2006).

[3] G. Jackeli and G. Khaliullin, Phys. Rev. Lett. **102**, 017205 (2009).

[4] S. K. Choi, R. Coldea, A. N. Kolmogorov, T. Lancaster, I. I. Mazin, S. J. Blundell, P. G. Radaelli, Y. Singh, P. Gegenwart, K. R. Choi, S. W. Cheong, P. J. Baker, C. Stock, and J. Taylor, Phys. Rev. Lett. **108**, 127204 (2012).

[5] H. S. Kim and H. Y. Kee, Phys. Rev. B **93**, 155143 (2016).

[6] S. Hwan Chun, J.-W. Kim, J. Kim, H. Zheng, C. C. Stoumpos, C. D. Malliakas, J. F. Mitchell, K. Mehlawat, Y. Singh, Y. Choi, T. Gog, A. Al-Zein, M. M. Sala, M. Krisch, J. Chaloupka, G. Jackeli, G. Khaliullin, and B. J. Kim, Nat. Phys. **11**, 462 (2015).

[7] J. Chaloupka, G. Jackeli, and G. Khaliullin, Phys. Rev. Lett. **105**, 027204 (2010).

[8] S. M. Winter, Y. Li, H. O. Jeschke, and R. Valentí, Phys. Rev. B **93**, 214431 (2016).

[9] P. P. Stavropoulos, D. Pereira, and H. Y. Kee, Phys. Rev. Lett. **123**, 037203 (2019).

[10] K. W. Plumb, J. P. Clancy, L. J. Sandilands, V. V. Shankar, Y. F. Hu, K. S. Burch, H. Y. Kee, and Y. J. Kim, Phys. Rev. B **90**, 041112(R) (2014).

[11] P. Czajka, T. Gao, M. Hirschberger, P. Lampen-Kelley, A. Banerjee, J. Yan, D. G. Mandrus, S. E. Nagler, and N. P. Ong, Nat. Phys. **17**, 915 (2021).

[12] A. Banerjee, P. Lampen-Kelley, J. Knolle, C. Balz, A. A. Aczel, B. Winn, Y. Liu, D. Pajerowski, J. Yan, C. A. Bridges, A. T. Savici, B. C. Chakoumakos, M. D. Lumsden, D. A. Tennant, R. Moessner, D. G. Mandrus, and S. E. Nagler, Npj Quantum Mater. **3**, 8 (2018).

[13] S. H. Do, S. Y. Park, J. Yoshitake, J. Nasu, Y. Motome, Y. S. Kwon, D. T. Adroja, D. J. Voneshen, K. Kim, T. H. Jang, J. H. Park, K. Y. Choi, and S. Ji, Nat. Phys. **13**, 1079 (2017).

[14] Y. Kasahara, T. Ohnishi, Y. Mizukami, O. Tanaka, S. Ma, K. Sugii, N. Kurita, H. Tanaka, J. Nasu, Y. Motome, T. Shibauchi, and Y. Matsuda, Nature **559**, 227 (2018).

[15] A. Banerjee, C. A. Bridges, J. Q. Yan, A. A. Aczel, L. Li, M. B. Stone, G. E. Granroth, M. D. Lumsden, Y. Yiu, J. Knolle, S. Bhattacharjee, D. L. Kovrizhin, R. Moessner, D. A. Tennant, D. G. Mandrus, and S. E. Nagler, Nat. Mater. **15**, 733 (2016).

[16] M. Majumder, M. Schmidt, H. Rosner, A. A. Tsirlin, H. Yasuoka, and M. Baenitz, Phys. Rev. B **91**, 180401 (2015).

[17] H. Takagi, T. Takayama, G. Jackeli, G. Khaliullin, and S. E. Nagler, Nat. Rev. Phys. **1**, 264 (2019).

[18] J. A. Sears, L. E. Chern, S. Kim, P. J. Bereciartua, S. Francoual, Y. B. Kim, and Y. J. Kim, Nat. Phys. **16**, 837 (2020).

[19] C. Hickey and S. Trebst, Nat. Commun. **10**, 530 (2019).

[20] C. Kim, H.-S. Kim, and J.-G. Park, J. Phys. Condens. Matter **34**, 023001 (2022).

[21] H. Liu, Int. J. Mod. Phys. B **35**, 2130006 (2021).

[22] A. Abragam and B. Bleaney, *Electron Paramagnetic Resonance of Transition Ions* (Oxford University, London, 1986).

[23] R. Sano, Y. Kato, and Y. Motome, Phys. Rev. B **97**, 014408 (2018).

[24] H. Liu and G. Khaliullin, Phys. Rev. B **97**, 014407 (2018).

[25] H. Liu, J. Chaloupka, and G. Khaliullin, Phys. Rev. Lett. **125**, 047201 (2020).

[26] M. Songvilay, J. Robert, S. Petit, J. A. Rodriguez-Rivera, W. D. Ratcliff, F. Damay, V. Balédent, M. Jiménez-Ruiz, P. Lejay, E. Pachoud, A. Hadj-Azzem, V. Simonet, and C. Stock, Phys. Rev. B **102**, 224429 (2020).

[27] C. Kim, J. Jeong, G. Lin, P. Park, T. Masuda, S. Asai, S. Itoh, H.-S. Kim, H. Zhou, J. Ma, and J.-G. Park, J. Phys. Condens. Matter **34**, 045802 (2022).

[28] W. Yao and Y. Li, Phys. Rev. B **101**, 085120 (2020).

[29] E. Lefrançois, M. Songvilay, J. Robert, G. Nataf, E. Jordan, L. Chaix, C. V. Colin, P. Lejay, A. Hadj-Azzem, R. Ballou, and V. Simonet, Phys. Rev. B **94**, 214416 (2016).

[30] W. Chen, X. Li, Z. Hu, Z. Hu, L. Yue, R. Sutarto, F. He, K. Iida, K. Kamazawa, W. Yu, X. Lin, and Y. Li, Phys. Rev. B **103**, L180404 (2021).

[31] X. Hong, M. Gillig, R. Hentrich, W. Yao, V. Kocsis, A. R. Witte, T. Schreiner, D. Baumann, N. Pérez, A. U. B. Wolter, Y. Li, B. Büchner, and C. Hess, Phys. Rev. B **104**, 144426 (2021).

[32] G. Lin, J. Jeong, C. Kim, Y. Wang, Q. Huang, T. Masuda, S. Asai, S. Itoh, G. Günther, M. Russina, Z. Lu, J. Sheng, L. Wang, J. Wang, G. Wang, Q. Ren, C. Xi, W. Tong, L. Ling, Z. Liu, L. Wu, J. Mei, Z. Qu, H. Zhou, X. Wang, J.-G. Park, Y. Wan, and J. Ma, Nat. Commun. **12**, 5559 (2021).

[33] C. H. Lee, S. Lee, Y. S. Choi, Z. H. Jang, R. Kalaivanan, R. Sankar, and K.-Y. Choi, Phys. Rev. B **103**, 214447 (2021).

[34] Y. Onose, T. Ideue, H. Katsura, Y. Shiomi, N. Nagaosa, and Y. Tokura, Science (80-. ). **329**, 297 (2010).

[35] T. Ideue, Y. Onose, H. Katsura, Y. Shiomi, S. Ishiwata, N. Nagaosa, and Y. Tokura, Phys. Rev. B **85**, 134411 (2012).

[36] M. Hirschberger, R. Chisnell, Y. S. Lee, and N. P. Ong, Phys. Rev. Lett. **115**, 106603 (2015).

[37] D. Watanabe, K. Sugii, M. Shimozawa, Y. Suzuki, T. Yajima, H. Ishikawa, Z. Hiroi, T. Shibauchi, Y. Matsuda, and M. Yamashita, Proc. Natl. Acad. Sci. **113**, 8653 (2016).

[38] Y. Kasahara, K. Sugii, T. Ohnishi, M. Shimozawa, M. Yamashita, N. Kurita, H. Tanaka, J. Nasu, Y. Motome, T. Shibauchi, and Y. Matsuda, Phys. Rev. Lett. **120**, 217205 (2018).

[39] M. Hirschberger, J. W. Krizan, R. J. Cava, and N. P. Ong, Science (80-. ). **348**, 106 (2015).

[40] H.-L. Kim, M. J. Coak, J. C. Baglo, K. Murphy, R. W. Hill, M. Sutherland, M. C. Hatnean, G. Balakrishnan, and J.-G. Park, Rev. Sci. Instrum. **90**, 103904 (2019).

[41] K. Sugii, M. Shimozawa, D. Watanabe, Y. Suzuki, M. Halim, M. Kimata, Y. Matsumoto, S. Nakatsuji, and M. Yamashita, Phys. Rev. Lett. **118**, 145902 (2017).

[42] H. Doki, M. Akazawa, H. Y. Lee, J. H. Han, K. Sugii, M. Shimozawa, N. Kawashima, M. Oda, H. Yoshida, and M. Yamashita, Phys. Rev. Lett. **121**, 97203 (2018).

[43] M. Akazawa, M. Shimozawa, S. Kittaka, T. Sakakibara, R. Okuma, Z. Hiroi, H.-Y. Lee, N. Kawashima, J. H. Han, and M. Yamashita, Phys. Rev. X **10**, 041059 (2020).

[44] C. A. Pocs, I. A. Leahy, H. Zheng, G. Cao, E.-S. Choi, S.-H. Do, K.-Y. Choi, B. Normand, and M. Lee, Phys. Rev. Res. **2**, 013059 (2020).

[45] See Supplemental materials, which includes Ref. [79-81], for a detailed discussion about reproducibility of the thermal Hall measurement, hysteresis in longitudinal thermal conductivity,

and calculation of magnon Hall conductivity.

[46] J. A. N. Bruin, R. R. Claus, Y. Matsumoto, N. Kurita, H. Tanaka, and H. Takagi, Nat. Phys. **18**, 401 (2022).

[47] Y. Yang, G.-M. Zhang, and F.-C. Zhang, Phys. Rev. Lett. **124**, 186602 (2020).

[48] T. Qin, J. Zhou, and J. Shi, Phys. Rev. B **86**, 104305 (2012).

[49] J. Cookmeyer and J. E. Moore, Phys. Rev. B **98**, 060412 (2018).

[50] A. M. Samarakoon, Q. Chen, H. Zhou, and V. O. Garlea, Phys. Rev. B **104**, 184415 (2021).

[51] W. Yao, K. Iida, K. Kamazawa, and Y. Li, arXiv:2203.00282 (2022).

[52] M. Yamashita, M. Akazawa, M. Shimozawa, T. Shibauchi, Y. Matsuda, H. Ishikawa, T. Yajima, Z. Hiroi, M. Oda, H. Yoshida, H.-Y. Lee, J. H. Han, and N. Kawashima, J. Phys. Condens. Matter **32**, 074001 (2020).

[53] H. Lee, J. H. Han, and P. A. Lee, Phys. Rev. B **91**, 125413 (2015).

[54] X. Li, B. Fauqué, Z. Zhu, and K. Behnia, Phys. Rev. Lett. **124**, 105901 (2020).

[55] T. Ideue, T. Kurumaji, S. Ishiwata, and Y. Tokura, Nat. Mater. **16**, 797 (2017).

[56] G. Grissonnanche, A. Legros, S. Badoux, E. Lefrançois, V. Zatko, M. Lizaire, F. Laliberté, A. Gourgout, J.-S. Zhou, S. Pyon, T. Takayama, H. Takagi, S. Ono, N. Doiron-Leyraud, and L. Taillefer, Nature **571**, 376 (2019).

[57] G. Grissonnanche, S. Thériault, A. Gourgout, M.-E. Boulanger, E. Lefrançois, A. Ataei, F. Laliberté, M. Dion, J.-S. Zhou, S. Pyon, T. Takayama, H. Takagi, N. Doiron-Leyraud, and L. Taillefer, Nat. Phys. **16**, 1108 (2020).

[58] M.-E. Boulanger, G. Grissonnanche, S. Badoux, A. Allaire, É. Lefrançois, A. Legros, A. Gourgout, M. Dion, C. H. Wang, X. H. Chen, R. Liang, W. N. Hardy, D. A. Bonn, and L. Taillefer, Nat. Commun. **11**, 5325 (2020).

[59] J.-Y. Chen, S. A. Kivelson, and X.-Q. Sun, Phys. Rev. Lett. **124**, 167601 (2020).

[60] S. Sim, H. Yang, H.-L. Kim, M. J. Coak, M. Itoh, Y. Noda, and J.-G. Park, Phys. Rev. Lett. **126**, 015901 (2021).

[61] S. Chaudhary, P. Srivastava, and S. Patnaik, in *AIP Conference Proceedings*, Vol. 1942 (2018), p. 130045.

[62] R. Takahashi and N. Nagaosa, Phys. Rev. Lett. **117**, 217205 (2016).

[63] G. Go, S. K. Kim, and K.-J. Lee, Phys. Rev. Lett. **123**, 237207 (2019).

[64] X. Zhang, Y. Zhang, S. Okamoto, and D. Xiao, Phys. Rev. Lett. **123**, 167202 (2019).

[65] H. Zhang, C. Xu, C. Carnahan, M. Sretenovic, N. Suri, D. Xiao, and X. Ke, Phys. Rev. Lett. **127**, 247202 (2021).

[66] M. Ye, G. B. Halász, L. Savary, and L. Balents, Phys. Rev. Lett. **121**, 147201 (2018).

[67] É. Lefrançois, G. Grissonnanche, J. Baglo, P. Lampen-Kelley, J. Yan, C. Balz, D. Mandrus, S. E. Nagler, S. Kim, Y.-J. Kim, N. Doiron-Leyraud, and L. Taillefer, arXiv:2111.05493 (2021).

[68] H. Li, T. T. Zhang, A. Said, G. Fabbris, D. G. Mazzone, J. Q. Yan, D. Mandrus, G. B. Halász, S. Okamoto, S. Murakami, M. P. M. Dean, H. N. Lee, and H. Miao, Nat. Commun. **12**, 3513 (2021).

[69] R. Hentrich, A. U. B. Wolter, X. Zotos, W. Brenig, D. Nowak, A. Isaeva, T. Doert, A. Banerjee,

P. Lampen-Kelley, D. G. Mandrus, S. E. Nagler, J. Sears, Y.-J. Kim, B. Büchner, and C. Hess, Phys. Rev. Lett. **120**, 117204 (2018).

[70] R. Hentrich, M. Roslova, A. Isaeva, T. Doert, W. Brenig, B. Büchner, and C. Hess, Phys. Rev. B **99**, 085136 (2019).

[71] Y. Kasahara, S. Suetsugu, T. Asaba, S. Kasahara, T. Shibauchi, N. Kurita, H. Tanaka, and Y. Matsuda, arXiv:2202.11947 (2022).

[72] S. Suetsugu, Y. Ukai, M. Shimomura, M. Kamimura, T. Asaba, Y. Kasahara, N. Kurita, H. Tanaka, T. Shibauchi, J. Nasu, Y. Motome, and Y. Matsuda, arXiv:2203.00275 (2022).

[73] O. Tanaka, Y. Mizukami, R. Harasawa, K. Hashimoto, K. Hwang, N. Kurita, H. Tanaka, S. Fujimoto, Y. Matsuda, E.-G. Moon, and T. Shibauchi, Nat. Phys. (2022).

[74] S.-S. Zhang, G. B. Halász, and C. D. Batista, Nat. Commun. **13**, 399 (2022).

[75] K. Hwang, A. Go, J. H. Seong, T. Shibauchi, and E.-G. Moon, Nat. Commun. **13**, 323 (2022).

[76] J. Nasu, J. Yoshitake, and Y. Motome, Phys. Rev. Lett. **119**, 127204 (2017).

[77] P. Czajka, T. Gao, M. Hirschberger, P. Lampen-Kelley, A. Banerjee, N. Quirk, D. G. Mandrus, S. E. Nagler, and N. P. Ong, arXiv:2201.07873 (2022).

[78] L. E. Chern, E. Z. Zhang, and Y. B. Kim, Phys. Rev. Lett. **126**, 147201 (2021).

[79] M. Hirschberger, PhD Thesis, Princet. Univ. Press (2017).

[80] T. Gao, PhD Thesis, Princet. Univ. Press (2020).

[81] R. Matsumoto and S. Murakami, Phys. Rev. Lett. **106**, 197202 (2011).

**Figures**

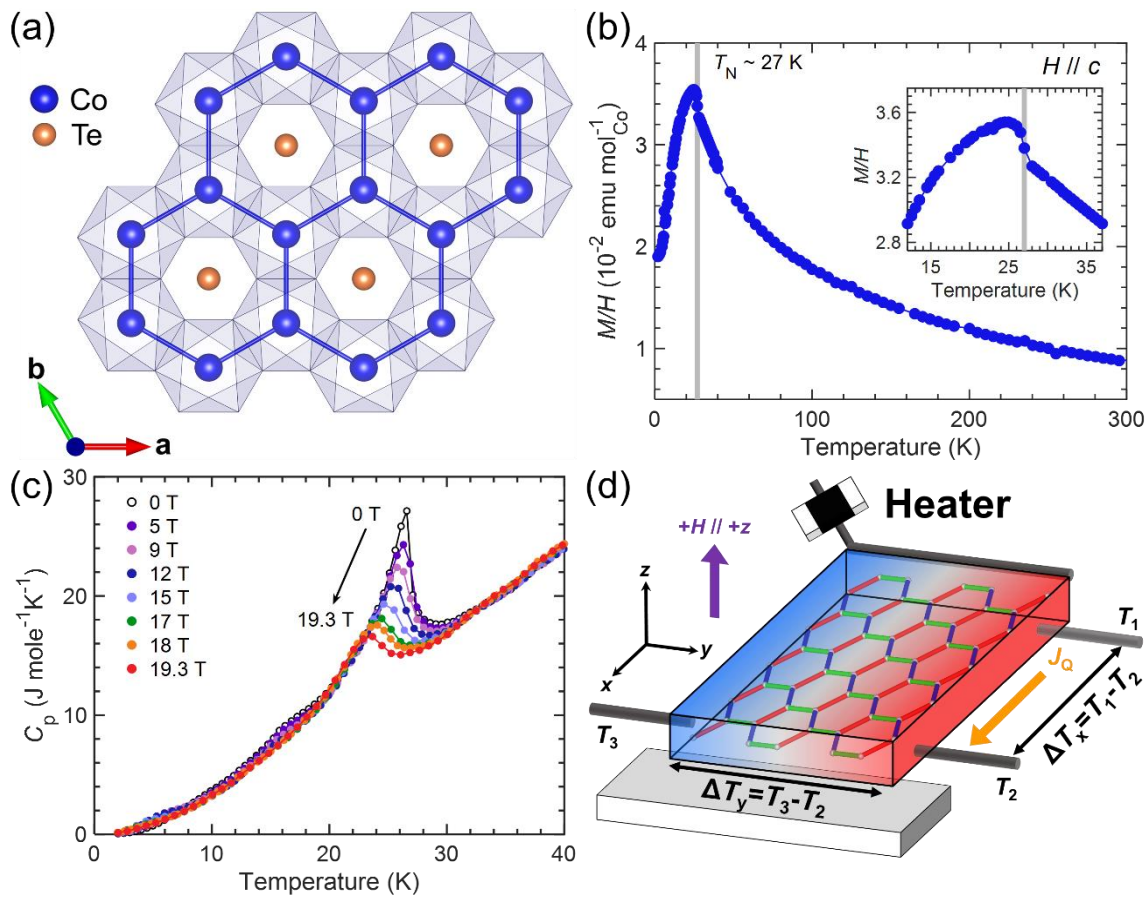

**Figure 1.** Crystallographic structure and thermodynamic measurement on $Na_2Co_2TeO_6$. (a) A crystallographic structure of Cobalt honeycomb layer in $Na_2Co_2TeO_6$, where edge-sharing $CoO_6$ octahedra form a honeycomb lattice with Na layer located between two Cobalt layers. (b) Temperature dependence of DC magnetic susceptibility (*M/H*) along the *c*-axis after field cooling under $\mu_0 H = 1$ T. The antiferromagnetic transition appears at $T_N \sim 27$ K. (c) Temperature-dependent heat capacity of $Na_2Co_2TeO_6$ under the various applied magnetic field along the *c*-axis up to 19.3 T. (d) Schematic of thermal transport experimental setup.

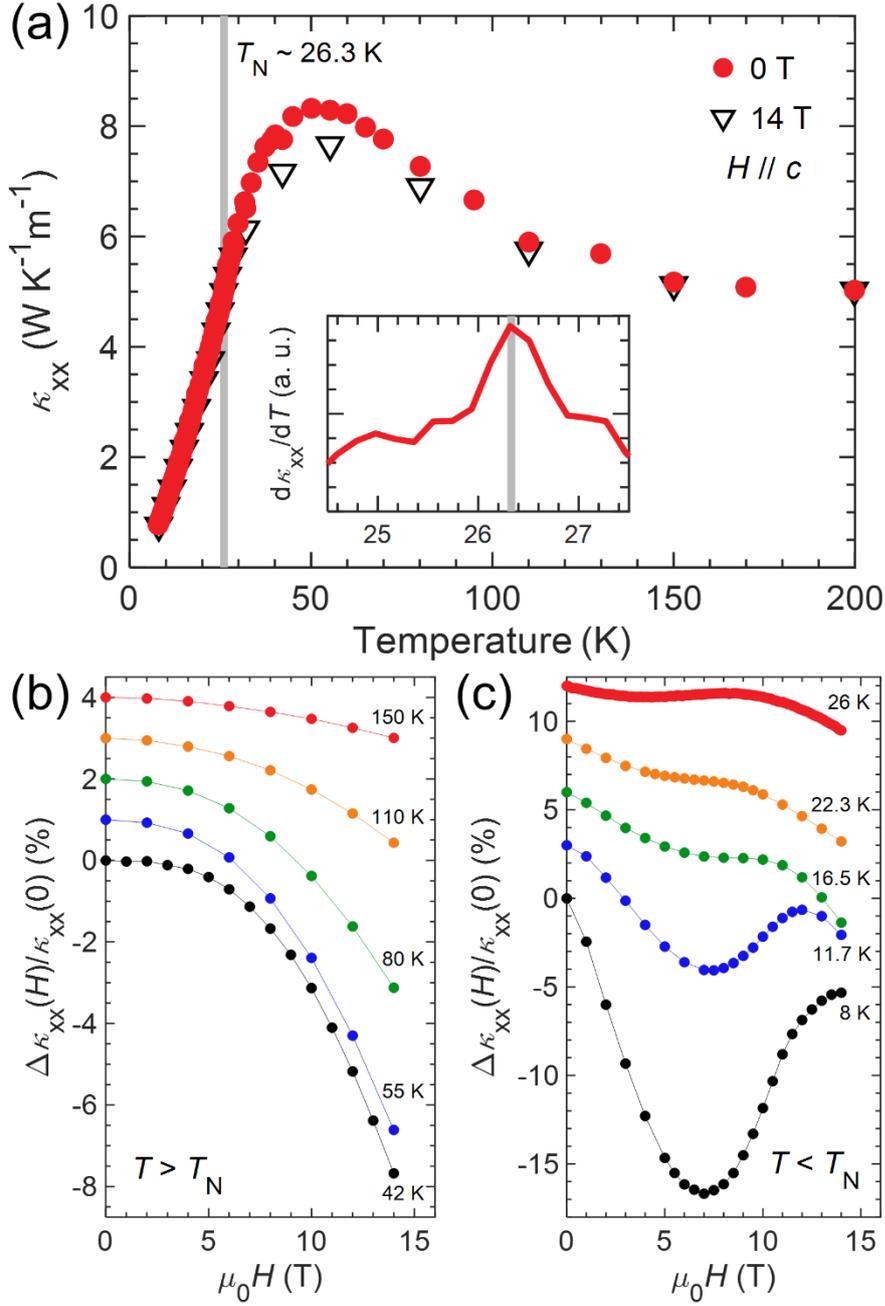

**Figure 2.** Longitudinal thermal conductivity ($\kappa_{xx}$) and magneto thermal conductivity $\Delta\kappa_{xx}(H)/\kappa_{xx}(0) \equiv [\kappa_{xx}(H) - \kappa_{xx}(0)]/\kappa_{xx}(0)$. (a) Temperature dependence of $\kappa_{xx}$. The heat current and magnetic field are applied along the *a*- and *c*-axes. (b) Magnetic field dependence of $\Delta\kappa_{xx}(H)/\kappa_{xx}(0)$ under the isothermal conditions (b) above and (c) below $T_N$. The data are shifted upwards along the *y*-axis for clarity.

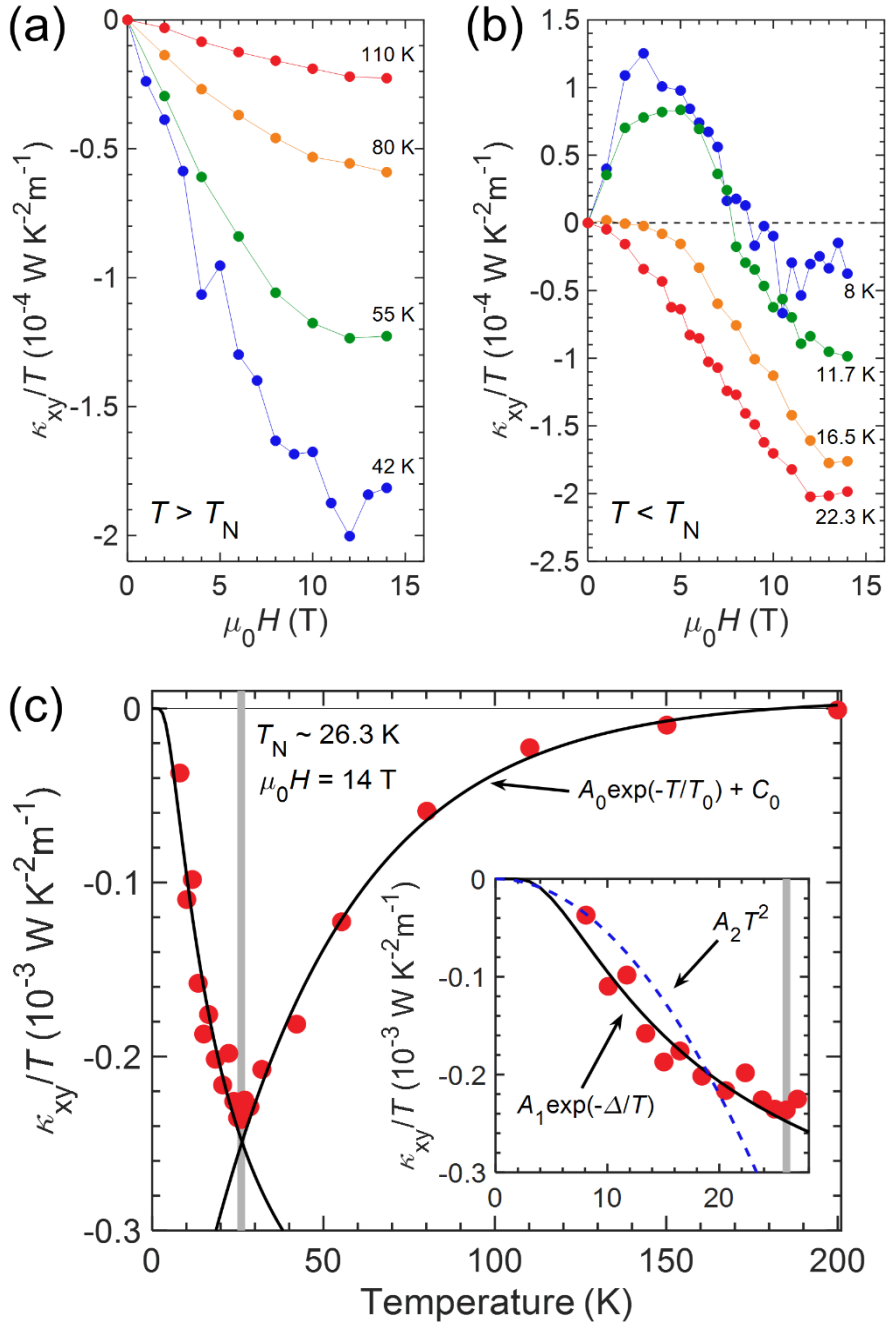

**Figure 3.** Thermal Hall conductivity divided by temperature ($\kappa_{xy}/T$). Magnetic field dependence of $\kappa_{xy}/T$ under the isothermal conditions (a) above and (b) below $T_N$. (c) $\kappa_{xy}/T$ as a function of temperature under the constant magnetic field of 14 T along the c-axis. The solid black curves are obtained using a recent scaling theory [47]. Data are fit to $A_0\exp(-T/T_0) + C_0$ with a fitting range above 32 K, and $A_1\exp(-\Delta/T)$ with a fitting range below 22 K, respectively. The inset is a blown-up picture showing more details about $\kappa_{xy}/T$ below $T_N$. The blue dashed curve is obtained from the fitting function $A_2T^2$. The fitting parameters are obtained as $A_0 = -0.48$ mW/K$^2$m, $A_1 = -0.45$ mW/K$^2$m, $A_2 = -5.5 \times 10^{-4}$ mW/K$^4$m, $C_0 = 0.0059$ mW/K$^2$m, $T_0 = 41.7$ K, $\Delta = 15.6$ K.

13